# Optical properties of cubic boron arsenide

Bai Song[1,2,3]*, Ke Chen[1]*, Kyle Bushick[4]*, Kelsey A. Mengle[4], Fei Tian[5], Geethal Amila Gamage[5], Zhifeng Ren[5†], Emmanouil Kioupakis[4†], Gang Chen[1†]

[1]*Department of Mechanical Engineering, Massachusetts Institute of Technology, Cambridge, MA 02139, USA*

[2]*Department of Energy and Resources Engineering, College of Engineering, Peking University, Beijing, China.*

[3]*Beijing Innovation Center for Engineering Science and Advanced Technology, Peking University, Beijing 100871, China*

[4]*Department of Materials Science and Engineering, University of Michigan, Ann Arbor, Michigan 48109, USA*

[5]*Department of Physics and Texas Center for Superconductivity, University of Houston, Houston, TX 77204, USA*

[†]Corresponding Authors. Email: zren2@central.uh.edu (Z.R.); kioup@umich.edu (E.K.); gchen2@mit.edu (G.C.)
*These authors contributed equally to this work.

**Abstract**

The ultrahigh thermal conductivity of boron arsenide makes it a promising material for next-generation electronics and optoelectronics. In this work, we report measured optical properties of cubic boron arsenide crystals including the complex dielectric function, refractive index, and absorption coefficient in the ultraviolet, visible, and near-infrared wavelength range. The data were collected at room temperature using spectroscopic ellipsometry as well as transmission and reflection spectroscopy. We further calculate the optical response using density functional and many-body perturbation theory, considering quasiparticle and excitonic corrections. The computed values for the direct and indirect band gaps (4.25 eV and 2.07 eV) agree well with the measured results (4.12 eV and 2.02 eV). Our findings contribute to the effort of using boron arsenide in novel electronic and optoelectronic applications that take advantage of its demonstrated ultrahigh thermal conductivity and predicted high ambipolar carrier mobility.

**1. Introduction**

Novel electronic and optoelectronic technologies often demand unconventional materials beyond silicon. In particular, III-V semiconductors such as gallium arsenide (GaAs), gallium nitride (GaN) and boron nitride (BN) are becoming increasingly important in applications ranging from efficient solar cells and solid-state lighting to high-power and high-speed transistors, largely thanks to their high carrier mobilities, wide band gaps, and high thermal conductivities. In the III-V family, cubic boron arsenide (BAs) has remained barely explored for decades because of challenges in crystal growth. However, very recently BAs has been drawing considerable attention due to the theoretical prediction and subsequent experimental demonstration of an ultrahigh isotropic thermal conductivity (~1200 Wm$^{-1}$K$^{-1}$) [1-4]. In addition, *ab initio* simulations predicted simultaneously high electron (~1400 cm$^2$V$^{-1}$s$^{-1}$) and hole mobilities (~2100 cm$^2$V$^{-1}$s$^{-1}$) in BAs [5]. These properties combined with an electronic bandgap around 1.5 eV to 2 eV [6-10] render BAs a promising material for novel electronic and optoelectronic applications.

Despite progress in the study of heat and charge transport in BAs, its optical properties remain largely unexplored and are the focus of the present work. Here we grow high-quality millimeter-sized BAs crystals and systematically measure the complex dielectric function in the ultraviolet, visible, and near-infrared spectral range using spectroscopic ellipsometry in conjunction with transmission/reflection spectroscopy. We further perform state-of-the-art computations based on density functional (DFT) and many-body perturbation theory with quasiparticle and excitonic corrections. The measured refractive index in the visible range lies within 3 to 3.5, comparable to many other III-V semiconductors and in good agreement with computed values. We observe strong absorption in good agreement with the calculations for photon energies higher than the indirect gap. The sub-band-gap absorption is attributed to surface and bulk defects and impurities that can potentially exist in BAs grown by chemical vapor transport technique. We also find that excitons have a sizeable effect on the direct absorption in BAs, which is consistent with previous estimates of exciton binding energies around 40 meV [6, 11]. Finally, we determine the experimental indirect band gap to be around 2.02 eV, and the experimental direct band gap to be around 4.12 eV. These values are in excellent agreement with the calculated values of 2.07 eV and 4.25 eV for the indirect and direct band gaps, respectively.

## 2. Experimental results

We grew high-quality millimeter-sized BAs crystals (Fig. 1a,b) using a seeded chemical vapor transport technique [12]. The sample thickness varies from a few micrometers for thin transparent platelets to hundreds of micrometers for opaque bulk samples. Excellent surface cleanliness, flatness, and smoothness are key to accurate sample characterizations, optical measurements in particular. We sequentially treated the as-grown samples with hydrochloric acid, acetone and isopropanol, followed by oxygen and argon plasma cleaning. The high crystal quality of the samples was verified by Raman spectroscopy (Fig. 1c, Horiba LabRAM HR Evolution) and single-crystal X-ray diffraction (Fig. 1d, Rigaku D-max IIIB X-Ray Diffractometer with a Cu K$\alpha$ radiation source). All the Raman and the XRD peak positions agree well with literature values [1-3]. The Raman peak around 699.5 cm$^{-1}$ shows a full width at half maximum (FWHM) of 1.1 cm$^{-1}$, and the rocking curve around the (111) peak shows a FWHM of 0.06°, indicating a relatively low level of crystal imperfections. Samples with flat, specularly reflecting surfaces were selected for further examination using atomic force microscopy (Veeco Dimension 3100). The root-mean-square surface roughness was measured to be 2 nm over a 50 μm × 50 μm scanning area (inset of Fig. 1b), good for the intended optical studies.

To characterize the optical response of the BAs crystals, we first employed spectroscopic ellipsometry [13]. Briefly, at any wavelength $\lambda$ of interest, the ratio $\rho$ between the complex Fresnel reflection coefficients for $p$- and $s$-polarized light ($r_p$ and $r_s$, respectively) is measured as

$$\rho = \frac{r_p}{r_s} = \tan\psi e^{i\Delta}, \tag{1}$$

where $\psi$ and $\Delta$ are the conventional ellipsometry angles with $\tan\psi$ capturing the amplitude ratio while $\Delta$ the phase difference. For a bulk opaque sample in air, an air/substrate (infinitely thick) model applies and the sample dielectric function $\varepsilon$ can be expressed as [13]

$$\varepsilon = \varepsilon_1 + i\varepsilon_2 = \sin^2\phi \left[1 + \left(\frac{1-\rho}{1+\rho}\right)^2 \tan^2\phi\right], \tag{2}$$

where $\varepsilon_1$ and $\varepsilon_2$ are respectively the real and imaginary part of the dielectric function, and $\phi$ is the incident angle. As shown in Eq. (2), both $\varepsilon_1$ and $\varepsilon_2$ are obtained simultaneously if the angles $\psi$ and $\Delta$ can be accurately measured. In addition, the complex refractive index $\tilde{n}$ is obtained as the square root of the dielectric function,

$$\tilde{n} = n + ik = \sqrt{\varepsilon}, \tag{3}$$

where $n$ and $k$ are the refractive index and extinction coefficient, respectively. With $n$ and $k$, the surface reflectivity and absorption coefficient $\alpha = 4\pi k/\lambda$ can be further calculated.

In order to measure the ellipsometry angles $\psi$ and $\Delta$, we employed a variable-angle rotating compensator ellipsometer (Woollam M-2000D) in the polarizer-compensator-sample-analyzer (PCSA) configuration. The wavelength range of 190 nm to 1000 nm (1.2 eV - 6.5 eV) was covered by using a quartz tungsten halogen lamp and a deuterium arc lamp as the sources.

For detection, a charge-coupled device (CCD) was used to allow fast data acquisition and reduced system drift. In order to measure millimeter-sized samples, focusing lenses were used to reduce the beam diameter to roughly 300 μm. The angular deviation from the nominal incident angle due to beam focusing is small and was accounted for during the fitting process using a proprietary software (Woollam WVASE32). After a careful system calibration, we first measured the ellipsometry angles on an intrinsic silicon (Si) wafer as a control experiment (Fig. 2a). The dielectric functions obtained are shown in Fig. 2b and agree well with literature values [14]. We then proceeded to characterize over ten opaque BAs samples, six of which were each measured at three randomly selected spots with a total of about ten measurements performed. Good agreement was observed among all samples. Long data acquisition time above one minute with a 20 Hz data acquisition rate was used to significantly increase the signal-to-noise ratio. In addition, we performed measurements at multiple incident angles including 65, 70, and 75 degrees to further improve the reliability of our results. The ellipsometry angles and the corresponding dielectric functions measured from a representative BAs sample (#c4) are shown in Fig. 2c and Fig. 2d, respectively. Although the ellipsometry angles at the three different incident angles largely differ, the extracted dielectric functions agree well.

Ellipsometry is highly sensitive to sample surface. In order to better capture the bulk properties, we further performed transmittance ($T$) and reflectance ($R$) measurements on BAs platelets about 10 μm thick. A homebuilt setup and a commercial UV-Vis-NIR spectrophotometer (Agilent Cary 5000) were used. In the homebuilt setup, the light source was an optical parametric amplifier (OPA, Light Conversion Orpheus). A plano-convex lens with 76 mm focal length was placed before the sample to focus the normal incident laser beam into a circular spot around 160 μm in diameter ($1/e^2$). The beam size on the lens is about 3 mm in diameter, thus the estimated maximum divergent angle of the focused beam is 1.1 deg, which is acceptably small enough not to induce significant measurement error due to focusing effects. A photodiode (Ophirs PD300, 5 nW-3 W) was used as the optical sensor to measure the transmitted and reflected laser powers at normal incidence. The wavelength range of 400 nm to 1100 nm was covered, as limited by the response of the photodiode. We measured BAs crystals about 10 μm thick and a few millimeters in lateral dimension, which were partially suspended over the edge of a glass slide. Based on the optical transfer matrix method [15], a simple thin-film-in-air model sufficed to describe the beam reflection and transmission, with the refractive index $n$ and the extinction coefficient $k$ of BAs as the only unknown parameters. Once $T$ and $R$ at each wavelength were measured experimentally, $n$ and $k$ were computed by solving the model. Measurements of $T$ and $R$ with the commercial spectrophotometer were performed in a similar way. The only difference is that the optical constants were extracted using an incoherent multi-optical-beams model [16] which accounts for the incoherent halogen light source in the spectrophotometer.

To verify the reliability of our homebuilt $T/R$ setup, we first characterized a 2 μm-thick Si membrane. Figure 3a shows the measured $T$ and $R$ data as a function of wavelength, together with the computed curves by using literature $n$ and $k$ values for Si [14]. The oscillations in the computed curves arise from interferences of the laser beams multiply reflected at the Si/air interfaces, especially when the wavelength becomes comparable to the membrane thickness. The measured data points are sparse due to the use of an OPA as the light source. Most of the measured data fall on the computed curves, although deviations are observed at a few points. Figure 3b shows the measured complex refractive index of Si, which agrees well with literature data [14]. Following the same procedure, two BAs platelets (sample #c2 and #c10) were then characterized using the homebuilt setup. In addition, a third BAs sample (#f1) was characterized using the commercial spectrophotometer. All the measured *T*/*R* data for BAs are shown in Fig. 3c. Figure 3d summarizes all the measured refractive index of BAs, along with our calculated results from DFT simulations. At wavelengths longer than 300 nm, the measured refractive index *n* approaches the computed curve and converges to a constant refractive index $n_0$ (3.04±0.02) in the near-infrared region. In the visible wavelength range, the extinction coefficient *k* from the *T*/*R* measurement and the DFT calculation are both very small (close to zero) when compared to *n* (see detailed comparison in Fig. 4b). The measured *k* from ellipsometry generally shows noticeable larger values than the DFT calculations, but is closer to the calculated result when exciton effect is considered at wavelengths shorter than 300 nm. The possible reason will be discussed later.

## 3. First-principles calculations

The procedure for calculating the quasiparticle band gaps and band structure of BAs is detailed in previous work [6]. We calculate the imaginary part of the dielectric function ($\varepsilon_2$) due to direct interband optical transitions [17] by interpolating the quasiparticle band structure and velocity matrix elements using the maximally localized Wannier function [18] method and the wannier90 code [19]. Fifty bands were included in the interpolation to converge the imaginary part of the dielectric function for photon energies ranging from 0-40 eV. The Brillouin zone (BZ) was sampled with a uniform 128×128×128 grid and the energy-conserving delta function was approximated with a Gaussian function with a broadening of 0.1 eV. Spin-orbit coupling corrections were not included in the calculations of the optical spectra as they reduce the band gap by only 66 meV [6], and do not appreciably change the calculated spectra, besides a rigid shift in the onset. We use the electron-phonon Wannier (EPW) code [20, 21] in order to calculate the imaginary dielectric function from phonon-assisted (indirect) optical transitions [22] from 0.05 eV to 5 eV at 300 K. We calculate phonon frequencies and displacements on a coarse 4×4×4 BZ-sampling grid and the electron quasiparticle energies and wave functions on an 8×8×8 BZ-sampling grid. The indirect $\varepsilon_2$ is calculated on finely interpolated BZ-sampling grids of 32×32×32 for the electron and 16×16×16 for the phonon wave vectors. The delta function was also approximated with a Gaussian of 0.1 eV broadening. We subsequently combine the direct and indirect contributions to the imaginary part of the dielectric function and employ the Kramers-Kronig relation to calculate the real part of the dielectric function ($\varepsilon_1$) in the 0-8 eV range. We then determine the frequency-dependent complex refractive index $\tilde{n} = n + ik$, according to Eq. (3); as well as the absorption coefficient $\alpha$, using $\alpha = 4\pi k/\lambda$. In order to examine excitonic effects on calculated properties, we utilize the Bethe-Salpeter equation method implemented in BerkeleyGW, using an 18×18×18 BZ sampling grid and a numerical broadening width of 0.2 eV. We calculate $\varepsilon_2$ up to 26 eV and use the same steps as above to calculate the other frequency dependent optical properties. Due to the small overlap of wave functions at the band extrema in an indirect-gap material such as silicon or BAs, excitonic effects are not expected to be significant in the indirect-absorption regime [22] and are not considered in this work. This expectation is further supported by the overall good agreement between our calculated phonon-assisted absorption spectra with the experimental measurements in the visible range, where indirect absorption is occurring.

## 4. Discussion

In Fig. 4a we present our calculated optical constants $n$ and $k$ of BAs as a function of photon energy, where it is evident that excitonic effects do appreciably modify these properties. The theoretical refractive index $n_0$ approaching zero photon energy is 2.99 without excitons, and 3.05 with excitons, agreeing excellently with the measured $n_0$ at long wavelength. We observe a steep increase in $k$ that coincides with the minimum direct band gap, and a peak around 6.4 eV, which we attribute to a large joint density of states at this energy [6]. In Fig. 4b, we plot the measured absorption coefficients as a function of photon energy (for wavelengths from 400 nm to 1100 nm) together with our calculated values. We note that our calculations have good qualitative agreement with other reports in the literature [23, 24], however, these previous works underestimate the band gap by about 0.5 eV, a point that Lyons et al. [11] also note in their recent work on BAs.

As we discuss above, the measured absorption is generally larger than calculated values. We propose two mechanisms to explain this disparity: defects/impurities and excitonic effects. Recently, TOF-SIMS and photoluminescence measurements have revealed that BAs crystals grown by chemical vapor transport (CVT) method typically possess a considerable amount of carbon, oxygen, and silicon defects and impurities [11]. Another work has recently used DFT to assess the thermodynamics of point defects and common impurities, finding that antisite pairs, $As_B$, and boron-related defects are the lowest energy native defects, while carbon impurities are also determined to be likely [25]. These findings support our hypothesis of strong defect absorption. Such impurities can form defect/impurity states within the band gap of BAs [11], providing additional channels for optical transitions, and hence increasing the absorption, especially at photon energies smaller than the indirect band gap (~2 eV, discussed later). With the existence of defects, the absorption coefficient measured from $T/R$ can be regarded as the summation of two parts: $\alpha_{T/R} = \alpha_{crystal} + \alpha_{defect}$. The absorption curve obtained from the calculation assumes a perfect crystalline system and only captures the $\alpha_{crystal}$ component of the "total" absorption. This pattern can be clearly viewed in Fig. 4c, where we plot the square root of $\alpha_{T/R}$ for three samples and compare with the calculation. From 1.1 eV to ~1.8 eV, instead of zero absorption as shown by the calculation, all $\alpha_{T/R}$ show a non-zero background, which we attribute to the defect absorption. In addition to revealing this background absorption, a plot of the square root of $\alpha$ vs. photon energy allows us to determine the indirect band gap value. The absorption coefficient associated with the indirect band gap transition can be expressed as [26]: $\alpha \propto \frac{(h\nu - E_{gi} + E_p)^2}{\exp(\frac{E_p}{kT}) - 1} + \frac{(h\nu - E_{gi} - E_p)^2}{-\exp(\frac{E_p}{kT}) + 1} \approx A(h\nu - E_{gi})^2$, where $h\nu$, $E_{gi}$, $E_p$, $T$, and $k$ are the photon energy, indirect band gap, energy of the phonon assisting in the transition, temperature, and the Boltzmann constant, respectively. This indicates that the square root of $\alpha$ should be linear with photon energy and the intersection is just the position of indirect band gap. In our case with additional defect absorption, the intersection is taken as the crossing point of the background and the increasing slope (which matches well with the slope of the calculated absorption), indicated by the dashed lines shown in Fig. 4c. The indirect band gaps determined in this way for our BAs samples are 1.98 eV(#f1), 2.03 eV(#c2), and 2.05 eV(#c10), close to the calculated value of 2.07 eV. As for the absorption coefficient measured with ellipsometry ($\alpha_{Ellips}$), the defect component $\alpha_{defect}$ overwhelms the intrinsic absorption $\alpha_{crystal}$ in the vicinity of indirect band gap (Fig. 4b), making $\alpha_{Ellips}$ inappropriate to yield indirect band gap information.

In addition to these defect and impurity contributions to absorption, we note that including excitonic effects is important for direct absorption. We find that for $\alpha_{Ellips}$ (which is very sensitive to surface conditions), $\alpha_{crystal}$ becomes comparable to $\alpha_{defect}$ in the UV regime (see $k$ in Fig. 3d), enabling the extraction of reliable intrinsic absorption information for short wavelengths. While defects may still contribute to absorption in the UV range, direct absorption is expected to dominate in this wavelength region. Additionally, as shown in Fig. 3d, the calculated absorption coefficient for direct transitions shows better agreement with experiment once excitonic effects are included. The fact that the slope of our calculated indirect absorption matches well with the measured absorption indicates that indirect excitons are weak. In Fig. 4d, we plot the square of $\alpha_{Ellips}$ whose measured range extends up to photon energies of ~6.5 eV, with the aim to determine the direct band gap of BAs. As the absorption associated with direct band gap transition can be expressed as [27]: $\alpha = A^*\sqrt{h\nu - E_{gd}}$, where $A^*$ and $E_{gd}$ are a material-related factor and the direct band gap, respectively. The square of $\alpha$ should be linear with the photon energy and the intersection is just the position of the direct band gap. The direct band gaps of our samples determined in this way are 4.09 eV (#c4) and 4.15 eV (#c5), close to the calculated value of 4.25 eV. Since the calculation is performed at 0 K, we do expect the measured value to be smaller at room temperature due to both zero-point motion effects and the temperature dependence of the band gap [28].

5. **Summary**

We studied the optical properties of single crystal BAs experimentally with ellipsometry and transmittance/reflectance measurements and theoretically using DFT and many-body perturbation theory including quasiparticle and excitonic corrections. In the visible range, the measured refractive index shows good agreement with the calculated results, with values varying from 3.0 to 3.5. The measured extinction (absorption) coefficients are in good agreement with calculated values, with the discrepancy in the sub-band-gap regime attributed to bulk and surface defect absorption. The measured indirect band gap (~2.02 eV) and the direct band gap (~4.12 eV) are consistent with the calculated values. Our results present the optical response of BAs crystals and provide a useful reference for the design of BAs-based photonics and optoelectronics.

**Acknowledgement**

This work was funded by the Multidisciplinary University Research Initiative (MURI) program, Office of Naval Research under a Grant No. N00014-16-1-2436 through the University of Texas at Austin (G.C. and Z.F. Ren). The computational work was supported by the Designing Materials to Revolutionize and Engineer our Future (DMREF) Program under Award No. 1534221, funded by the National Science Foundation (E.K.). K.A.M. acknowledges the support from the National Science Foundation Graduate Research Fellowship Program through Grant No. DGE 1256260. This research used resources of the National Energy Research Scientific Computing Center, a DOE Office of Science User Facility supported by the Office of Science of the U.S. Department of Energy under Contract No. DE-AC02-05CH11231

**References**

1. Kang, J. S.; Li, M.; Wu, H.; Nguyen, H.; Hu, Y., Experimental observation of high thermal conductivity in boron arsenide. *Science* 2018, 361, 575.
2. Li, S.; Zheng, Q.; Lv, Y.; Liu, X.; Wang, X.; Huang, P. Y.; Cahill, D. G.; Lv, B., High thermal conductivity in cubic boron arsenide crystals. *Science* 2018, 361, 579.
3. Tian, F.; Song, B.; Chen, X.; Ravichandran, N. K.; Lv, Y.; Chen, K.; Sullivan, S.; Kim, J.; Zhou, Y.; Liu, T.-H.; Goni, M.; Ding, Z.; Sun, J.; Udalamatta Gamage, G. A. G.; Sun, H.; Ziyaee, H.; Huyan, S.; Deng, L.; Zhou, J.; Schmidt, A. J.; Chen, S.; Chu, C.-W.; Huang, P. Y.; Broido, D.; Shi, L.; Chen, G.; Ren, Z., Unusual high thermal conductivity in boron arsenide bulk crystals. *Science* 2018, 361, 582.
4. Lindsay, L.; Broido, D. A.; Reinecke, T. L., First-Principles Determination of Ultrahigh Thermal Conductivity of Boron Arsenide: A Competitor for Diamond? *Physical Review Letters* 2013, 111, 025901.
5. Liu, T.-H.; Song, B.; Meroueh, L.; Ding, Z.; Song, Q.; Zhou, J.; Li, M.; Chen, G., Simultaneously high electron and hole mobilities in cubic boron-V compounds: BP, BAs, and BSb. *Physical Review B* 2018, 98, 081203.
6. Bushick, K.; Mengle, K.; Sanders, N.; Kioupakis, E., Band structure and carrier effective masses of boron arsenide: Effects of quasiparticle and spin-orbit coupling corrections. *Applied Physics Letters* 2019, 114, 022101.
7. Buckeridge, J.; Scanlon, D. O., Electronic band structure and optical properties of boron arsenide. *Physical Review Materials* 2019, 3, 051601.

8. Hart, G. L. W.; Zunger, A., Electronic structure of BAs and boride III-V alloys. *Physical Review B* 2000, 62, 13522-13537.
9. Surh, M. P.; Louie, S. G.; Cohen, M. L., Quasiparticle energies for cubic BN, BP, and BAs. *Physical Review B* 1991, 43, 9126-9132.
10. Wang, S.; Swingle, S. F.; Ye, H.; Fan, F.-R. F.; Cowley, A. H.; Bard, A. J., Synthesis and Characterization of a p-Type Boron Arsenide Photoelectrode. *Journal of the American Chemical Society* 2012, 134, 11056-11059.
11. Lyons, J. L.; Varley, J. B.; Glaser, E. R.; Freitas, J. A.; Culbertson, J. C.; Tian, F.; Gamage, G. A.; Sun, H.; Ziyaee, H.; Ren, Z., Impurity-derived p-type conductivity in cubic boron arsenide. *Applied Physics Letters* 2018, 113, 251902.
12. Tian, F.; Song, B.; Lv, B.; Sun, J.; Huyan, S.; Wu, Q.; Mao, J.; Ni, Y.; Ding, Z.; Huberman, S.; Liu, T.-H.; Chen, G.; Chen, S.; Chu, C.-W.; Ren, Z., Seeded growth of boron arsenide single crystals with high thermal conductivity. *Applied Physics Letters* 2018, 112, 031903.
13. Woollam, J. A.; Johs, B. D.; Herzinger, C. M.; Hilfiker, J. N.; Synowicki, R. A.; Bungay, C. L., *Overview of variable-angle spectroscopic ellipsometry (VASE): I. Basic theory and typical applications*. SPIE: 1999; Vol. 10294.
14. Green, M. A., Self-consistent optical parameters of intrinsic silicon at 300K including temperature coefficients. *Solar Energy Materials and Solar Cells* 2008, 92, 1305-1310.
15. Katsidis, C. C.; Siapkas, D. I., General transfer-matrix method for optical multilayer systems with coherent, partially coherent, and incoherent interference. *Appl. Opt.* 2002, 41, 3978-3987.
16. Schinke, C.; Christian Peest, P.; Schmidt, J.; Brendel, R.; Bothe, K.; Vogt, M. R.; Kröger, I.; Winter, S.; Schirmacher, A.; Lim, S.; Nguyen, H. T.; MacDonald, D., Uncertainty analysis for the coefficient of band-to-band absorption of crystalline silicon. *AIP Advances* 2015, 5, 067168.
17. Rondinelli, J. M.; Kioupakis, E., Predicting and Designing Optical Properties of Inorganic Materials. *Annual Review of Materials Research* 2015, 45, 491-518.
18. Marzari, N.; Mostofi, A. A.; Yates, J. R.; Souza, I.; Vanderbilt, D., Maximally localized Wannier functions: Theory and applications. *Reviews of Modern Physics* 2012, 84, 1419-1475.
19. Mostofi, A. A.; Yates, J. R.; Lee, Y.-S.; Souza, I.; Vanderbilt, D.; Marzari, N., wannier90: A tool for obtaining maximally-localised Wannier functions. *Computer Physics Communications* 2008, 178, 685-699.
20. Poncé, S.; Margine, E. R.; Verdi, C.; Giustino, F., EPW: Electron–phonon coupling, transport and superconducting properties using maximally localized Wannier functions. *Computer Physics Communications* 2016, 209, 116-133.
21. Giustino, F.; Cohen, M. L.; Louie, S. G., Electron-phonon interaction using Wannier functions. *Physical Review B* 2007, 76, 165108.
22. Noffsinger, J.; Kioupakis, E.; Van de Walle, C. G.; Louie, S. G.; Cohen, M. L., Phonon-Assisted Optical Absorption in Silicon from First Principles. *Physical Review Letters* 2012, 108, 167402.
23. Bravić, I.; Monserrat, B., Finite temperature optoelectronic properties of BAs from first principles. *Physical Review Materials* 2019, 3, 065402.
24. Ge, Y.; Wan, W.; Guo, X.; Liu, Y., The direct and indirect optical absorptions of cubic BAs and BSb. *eprint arXiv:1901.03947* 2019, arXiv:1901.03947.
25. Chae, S.; Mengle, K.; Heron, J. T.; Kioupakis, E., Point defects and dopants of boron arsenide from first-principles calculations: Donor compensation and doping asymmetry. *Applied Physics Letters* 2018, 113, 212101.
26. Pankove, J. I., *Optical processes in semiconductors*. Courier Corporation: 1975.
27. Rosencher, E.; Vinter, B., *Optoelectronics*. Cambridge University Press: 2002.
28. Giustino, F.; Louie, S. G.; Cohen, M. L., Electron-phonon renormalization of the direct band gap of diamond. *Physical review letters* 2010, 105, 265501.

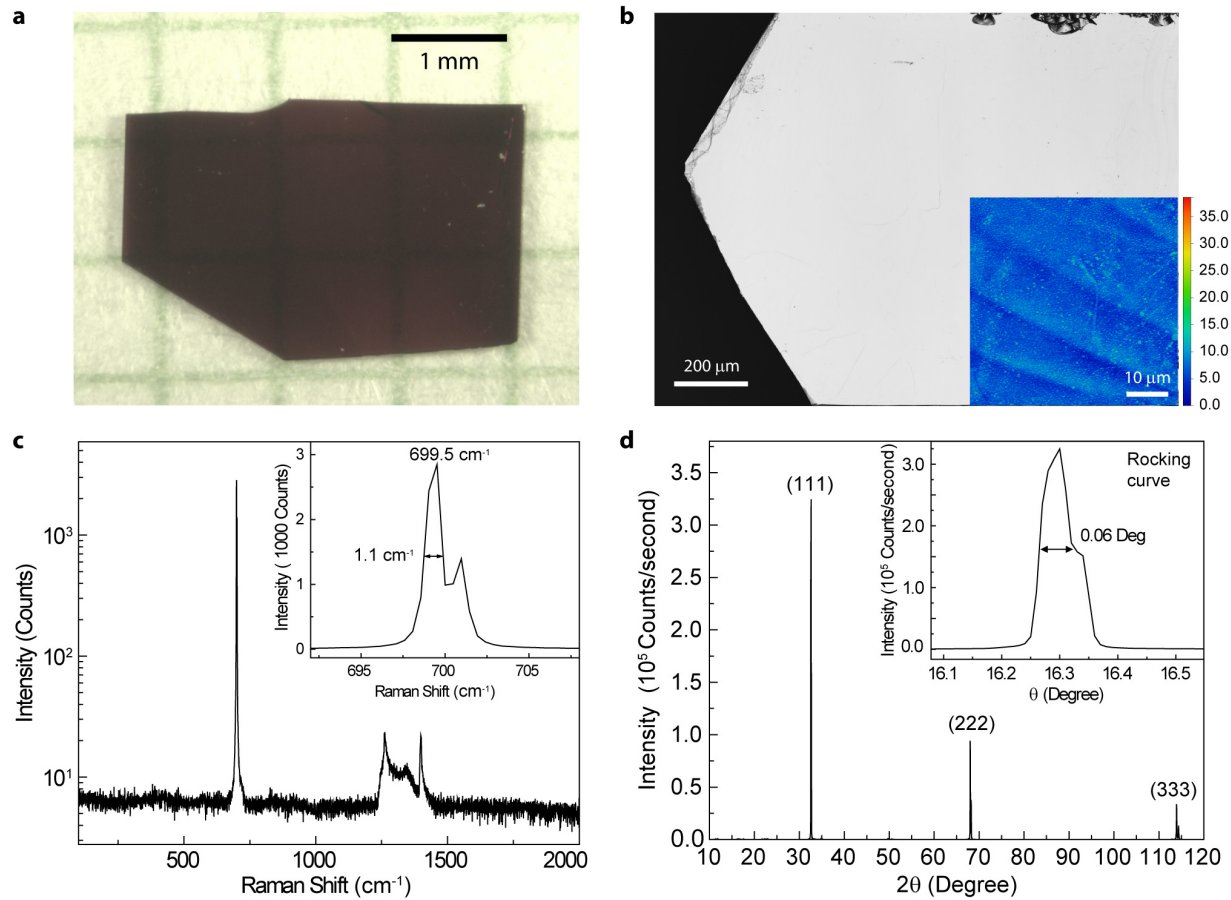

Figure 1. Characterization of the boron arsenide crystals. (a) Image of a BAs sample of moderate thickness under an optical microscope. (b) Laser confocal scanning microscopy image of a BAs sample showing a smooth and clean surface. The inset is a large-area atomic force microscopy image which yields a root-mean square surface roughness of 2 nm. (c) A representative Raman spectrum. The inset zooms in around the main peak and shows a full width at half maximum (FWHM) of 1.1 cm$^{-1}$. (d) Single-crystal X-ray diffraction pattern of representative BAs sample. The rocking curve around the (111) peak (inset) shows a FWHM of 0.06°.

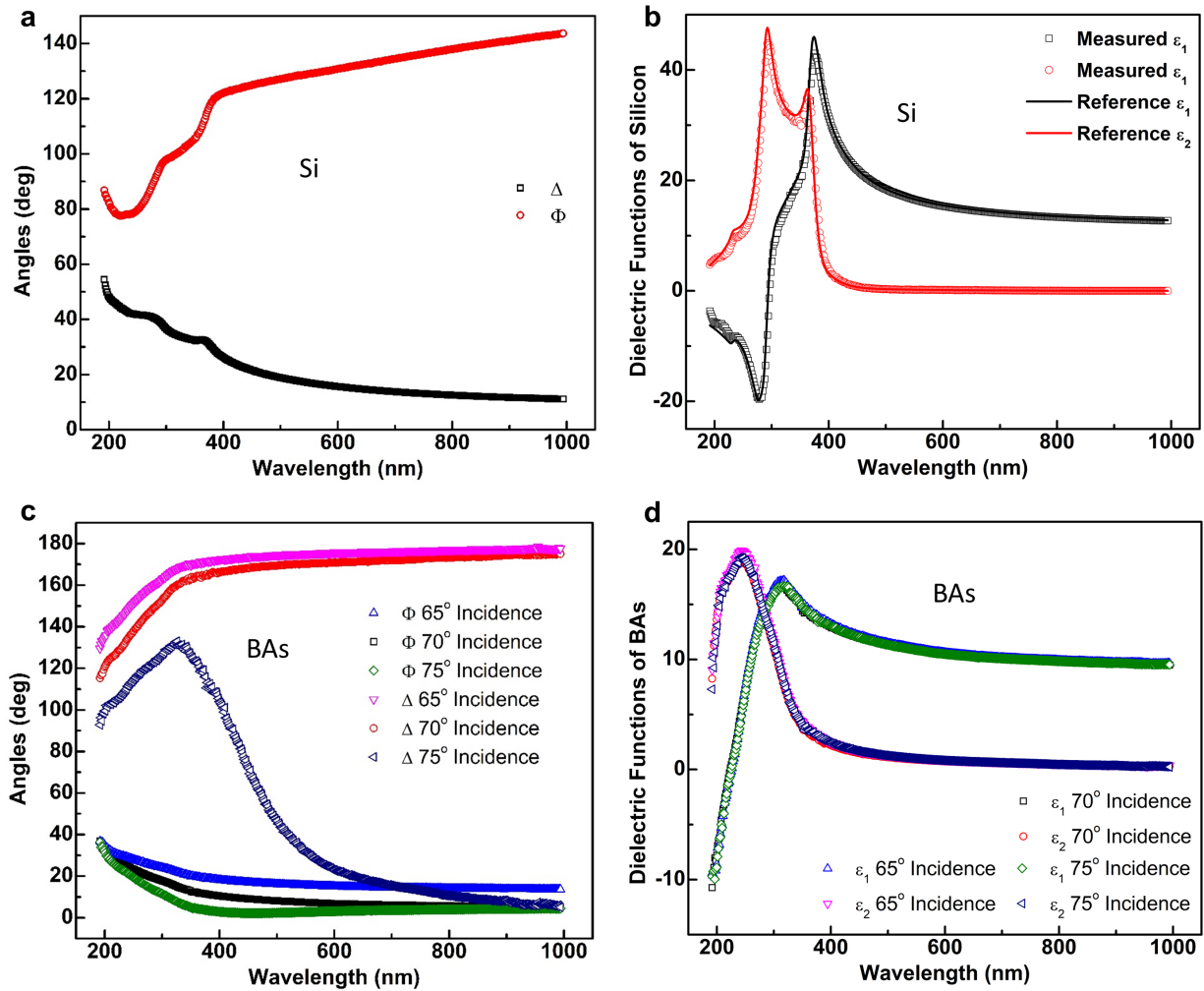

Figure 2. (a) Ellipsometry angles measured on an intrinsic silicon wafer. (b) The dielectric function of silicon obtained from the ellipsometry data, along with reference values. (c) Representative ellipsometry data from a thick BAs crystal, measured at three different incident angles. (d) The corresponding dielectric functions.

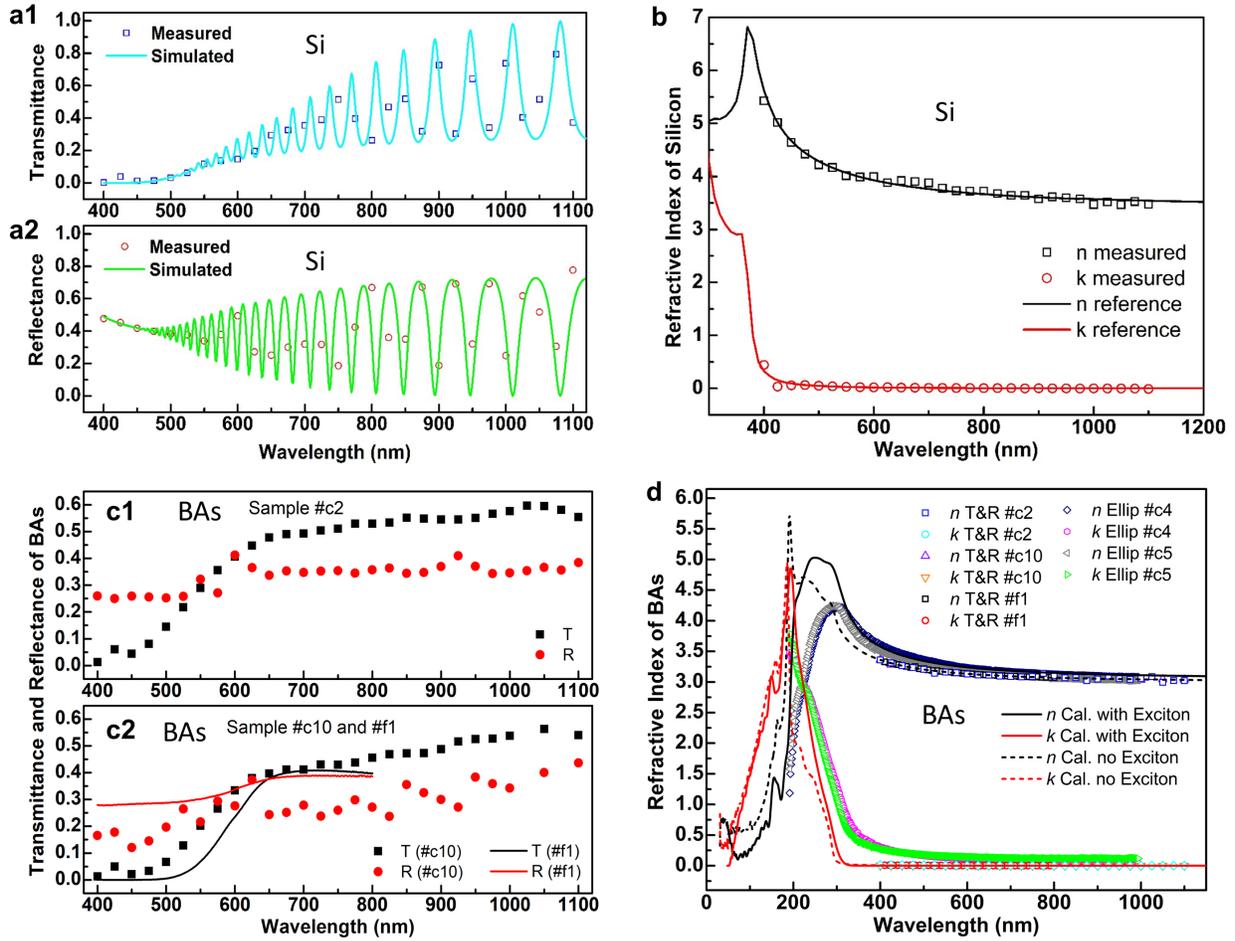

Figure 3. (a) Measured transmittance (a1) and reflectance (a2) of a 2 μm-thick silicon membrane, along with simulated transmittance using reference refractive index from the literature. (b) Measured complex refractive index of the silicon membrane from the transmittance and reflectance data, together with the reference data. (c) Measured $T$ and $R$ of thin BAs flakes from sample #c2 (c1) and sample #c10 (c2), respectively. Measured $T$ and $R$ with the commercial spectrophotometer from BAs sample #f1 are also shown in (c2). (d) Refractive index of BAs, from the ellipsometry and the $T$/$R$ measurements, and the DFT calculation.

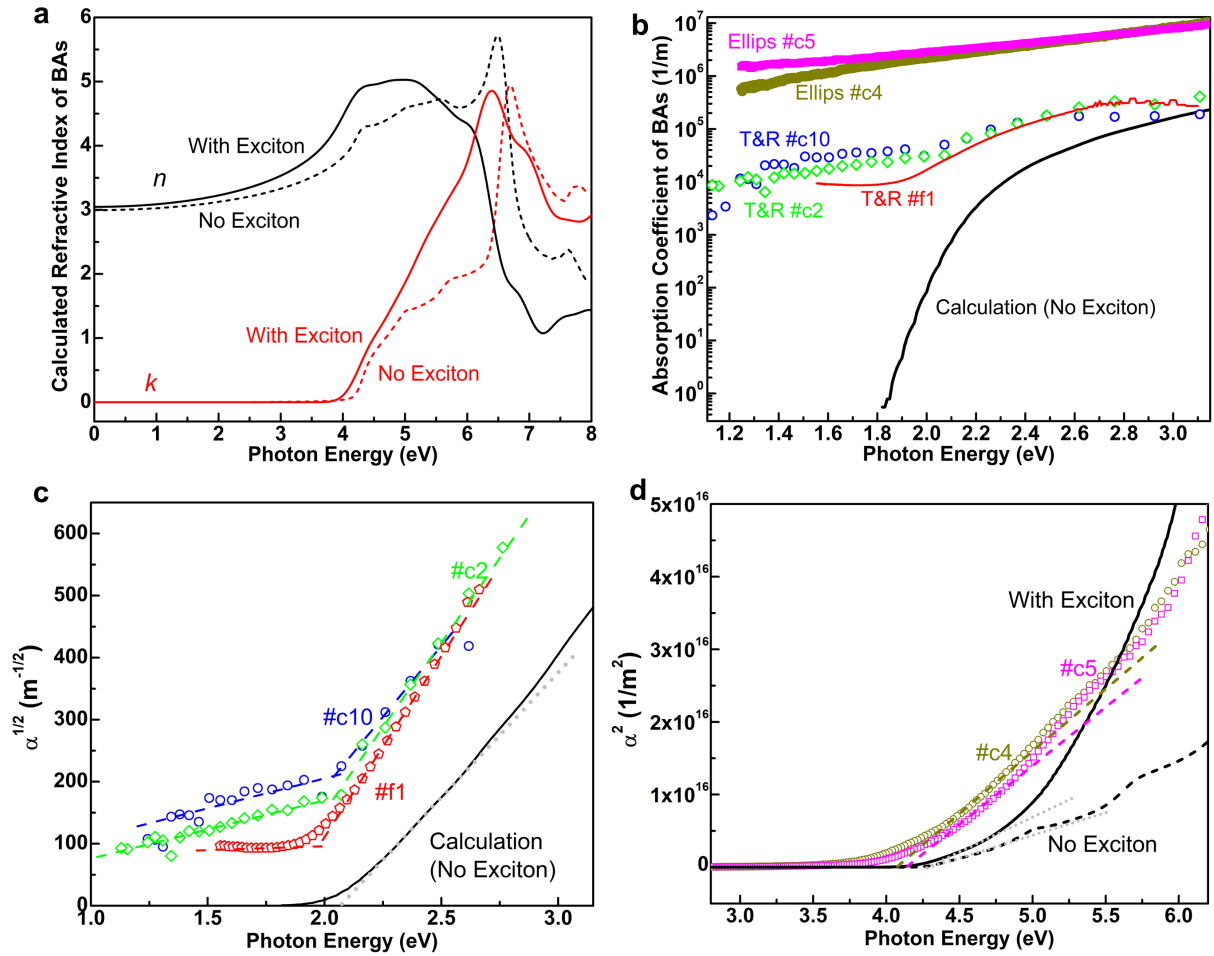

Figure 4. (a) Calculated complex refractive index of BAs as a function of photon energy. (b) Comparison of absorption coefficients of BAs from ellipsometry, *T*/*R* measurements, and the DFT calculation. (c) Square root of absorption coefficient of BAs from *T*/*R* measurements and DFT calculation. (d) Square of absorption coefficient of BAs from ellipsometry and DFT calculation. Color dashed lines and grey dotted lines are given as visual guides within (c) and (d).

19